\documentclass {article}
\usepackage[utf8]{inputenc}
\usepackage[english]{babel}
\usepackage {amssymb,amsmath,epsfig}

\title{Data Load Balancing in Heterogeneous Dynamic Networks}
\author{Iman Mirrezaei\\
	Sharif University of Technology \\ imirrezaei@alum.shairf.edu \\
	\and 
	Javad Shahparian \\
	Sharif University of Technology \\ jshahparian@alum.shairf.edu \\}
\date{ }
\begin {document}
\maketitle
\begin {abstract}
  Data load balancing is a challenging task in the P2P systems.
 Distributed hash table (DHT) abstraction, heterogeneous nodes, and non uniform distribution of objects are the reasons to cause load imbalance in structured P2P overlay networks.
 Previous works solved the load balancing problem by assuming  the homogeneous capabilities of nodes, unawareness of the link latency during transferring load, and imposing logical structures to collect and reassign load. 
 We propose a distributed load balancing algorithm with the topology awareness  feature by using the concept of virtual servers. 
 In our approach, each node collects neighborhood load information from physically close  nodes  and  reassigns  virtual servers to overlay nodes based upon the topology of underlying network.
 Consequently, our approach converges data load balancing quickly and it also reduces the load transfer cost between nodes.
 Moreover, our approach increases the quality of load balancing  among  close nodes of overlay and it also introduces a new tradeoff between the quality of load balancing and load transfer cost among all overlay nodes. 
 Our simulations show that our approach reduces the load transfer cost and it saves network bandwidth respectively.
 Finally, we show that the in-degree imbalance of nodes, as a consequence of  topology awareness, cannot lead to a remarkable problem in topology aware overlays.
\end {abstract}

\section{Introduction} \label{Inroduction}
Structured P2P overlay networks~\cite{pastry,CAN,chord} provide  DHT abstraction for object storage and retrieval. In these overlays, each object  and node is identified by a  unique identifier. The search space is partitioned among overlay nodes and each node is responsible for storage and retrieval of objects in its region. These systems assume that resources such as network bandwidth, capacity and  storage are uniformly distributed among all participants of network.

In these structured systems, a distributed hash function chooses identifier of nodes and objects. So, it causes an $O(\log N)$ imbalance factor in the number of stored objects at a node. Moreover, if identifier of nodes are no longer uniformly distributed, the imbalance factor  becomes worse. This could happen in database applications because all data items (tuples) of a relation are kept with regard to  their primary key values(identifiers). In addition, load imbalance becomes worse  when there exist many nodes with different capabilities (storage, bandwidth, CPU, etc.). Resulted load imbalance deteriorates the functionality of overlay networks.

Several solutions have been proposed to resolve the load balancing problem~\cite{chord,disloadbalancing,loadbalancesp,proximity-Awareload,loadbalancedsp,CFS,ResourceSharing,kargarLoadbalancing}. But, these extensions have their own restrictions. Firstly, it may be assumed that nodes have similar capabilities. Secondly,  they normally ignore the link latency between nodes and extra load of a node may traverse the high link latency, thereby increasing the bandwidth consumption, increasing traffic in underlying network, and delaying the convergence of load balancing. Thirdly, these solutions may use some logical fixed nodes to collect load information and plan new reassignments. However, these solutions reduce the load balancing problem to a centralized problem, therefore, leading to the single point of failure problem and limited scalability.

This paper presents a distributed load balancing algorithm with topology awareness in which the mentioned restrictions are dealt with.
 We extended our previous approach~\cite{LoadBalancingRAQNet} for data load balancing in overlay networks.
Our algorithm  uses the concept of \emph {virtual servers}, formerly suggested in Chord~\cite{chord}: each node collects neighborhood load information from  \emph{close}  nodes according to the topology of the
underlying network  and reassigns  load. Then, virtual servers are transferred between physically  \emph{close} nodes.
It consequently  provides a rapid load balance convergence, replies quickly  to load imbalance,
reduces the load transfer cost, and improves the load balancing traffic.
This approach, while collecting load  information of nodes, does not impose any logical structure or overhead on the overlay network.
Moreover, it  does well in terms of scalability.
Our parametric algorithm  increases the quality of load balancing among   \emph{close} nodes of the overlay and also  provides a different kind of tradeoff between  the quality of load balancing and load transfer cost across overlay nodes.
In addition, each node or group of nodes can perform the proposed load balancing algorithm based on its
desired network distance. We perform our load balancing algorithm on the RAQNet~\cite{RAQNet} overlay network. In RAQNet overlay network, each node has a practical internet coordinate  (\emph{PIC})   for estimating internet network distances  between nodes by the \emph{PIC} mechanism~\cite{PIC}.

This solution can be performed in other structured  P2P overlay networks, if each overlay node knows its  practical internet coordinate~\cite{PIC}.
Particularly, we make the following contributions:
\begin{enumerate}
    \item We offer a fully distributed load balancing algorithm which transfers extra load   between physically  \emph{close} nodes based on the topology of the underlying network.
\item We use a simple parametric algorithm to collect load information of \emph{close} nodes, reassign load and provide  a practical tradeoff between load transfer cost, and the quality of load among all overlay nodes.
\item Our experimental results show that the network bandwidth is greatly saved and the load transfer cost is reduced in our approach.
\end{enumerate}

Furthermore, we consider how to maintain the topology aware property affects on the in-degree of nodes in the RAQNet overlay. The in-degree balance is an attractive feature in overlays because it balances the query routing. We conclude that this imbalance does not cause a significant problem.
In section~\ref{nodelinks},  we propose a reactive approach for it.
The rest of this paper is organized as follows.
Section \ref{relatedwork} provides a survey of the related work.
A brief overview of RAQNet overlay network are presented in section \ref{RAQNETSec}. Section \ref{topologyloadsec} describes the topology aware load balancing algorithm. The experimental evaluations are presented in section \ref{Exresult} and section \ref{Concsec} concludes the paper.

\section{Related work}\label{relatedwork}
Load balancing~\cite{disloadbalancing,loadbalancesp,proximity-Awareload,loadbalancedsp} is a significant and challenging step through data integration~\cite{OAEI,Consolidation,ConreferenceResolution} in distributed heterogeneous resources. An effective load balancing approach is required to achieve a scalable data integration process and utilize available nodes equally~\cite{Kolb}.
Most Structured P2P systems~\cite{pastry,CAN,chord} suppose that object IDs are distributed by the uniform hash function. Additionally, they suppose that all nodes have similar capacities and load. Even so, the resulted load balance is not completely perfect and they have an $O(\log n)$ imbalance load.

Many load balancing methods have been suggested to handle this problem in P2P systems. The first work has been done by Chord~\cite{chord}. They diminish  load of overlay nodes by using the concept of virtual servers. They allocate $\log N$ virtual servers per physical node and suppose that all overlay nodes are similar. However, their approach does not practically resolve the load balancing problem.

CFS~\cite{CFS} does not ignore the heterogeneity of nodes. In CFS, virtual servers are allocated to nodes according to their capacities. Also, they use a simple solution to transfer extra load from heavy nodes,
but their method may cause other nodes become overloaded.

Triantafillou et al.~\cite{ResourceSharing}  introduce the novel design to perform fair load distribution in the context of content and resource management in unstructured P2P systems. They collect load objects
by the meta-data and after that they compute a reassignment of objects by using that information.

Karger and Ruhl~\cite{kargarLoadbalancing} present dynamic load balancing algorithms without using virtual servers. In their algorithms, lightly loaded nodes should be neighbors of heavily loaded nodes in order to reassign their load. They maximize utilization of load in nodes but they do not completely consider
different node capacities. Moreover, It is not clear whether their algorithms are
 practical or not.

Roa et al.~\cite{loadbalancesp} propose three simple load balancing algorithms for DHT-based  systems: \emph{one-to-one, one-to-many} and \emph{many-to-many}. They transfer load from heavy nodes to light
 nodes in every unit of virtual servers. In their load balancing approach, they use \emph{directory} nodes to store load information of nodes and reassign virtual servers. \emph{one-to-many} and \emph{many-to-many} are extended by Godfrey et al.~\cite{loadbalancedsp}to perform load balancing in dynamic P2P systems. Their results   have shown that their approach is so effective, but they have two weak points. Firstly, their approach suffers from a single point of failure problem because of using  \emph{directory} nodes. Secondly, their approach does not notice to link latency between light and heavy nodes while transferring load.

Yingwu et al.~\cite{proximity-Awareload} use a \emph{k-ary} tree and virtual servers to perform load balancing in structured overlay networks.  In their algorithm, the load information is collected by the \emph{k-ary} tree and reassignments of virtual servers are scheduled by  nodes of the \emph{k-ary} tree. They use landmark binning~\cite{Topologically-aware-overlay} to manage virtual server assignments across nodes which are
  \emph{close} to each other according to topology of underlying network. They determine \emph{close} nodes by measuring from the landmark sites. Thus, landmark sites become hot spots  while the P2P system size are increasing.

The topology-aware load balancing algorithm presented in this paper is similar to
a distributed load balancing algorithm  proposed by Zhenyu et al~\cite{disloadbalancing}. In both algorithms, load are transferred based on topology information, but we collect load information of \emph{close}
 nodes by a restricted flooding algorithm with regard  to topology of underlying network. Our approach
improves the load balancing traffic and also provides rapid convergence on load balance. Also, our parametric algorithm increases the quality of load balancing among \emph{close} nodes of overlay and also  provides a different kind of tradeoff between the quality of load balancing and load transfer cost across all overlay nodes.
Moreover, each node or group of nodes can perform the proposed load balancing algorithm based on its desired network distance to transfer extra load.
\section{Overview of RAQNet}\label{RAQNETSec}
RAQNet~\cite{RAQNet}  is a multi-dimensional topology-aware overlay network  based on RAQ~\cite{RAQ} data structure. In RAQNet overlay network, the search space is $d$-dimensional Cartesian coordinate space which is
partitioned among $n$ nodes of the overlay network by a partition
tree. Each node has $O(\log n)$ links to other nodes. Each single
point query is routed via $O(\log n)$ message passing. Each node is corresponded to a region and it is responsible for  the queries targeting any point in its region.
In RAQNet overlay, nodes are connected to each other if they have the
same labels and also are \emph{close} to each other with respect to the topology of
the underlying network. A topological match between an overlay and its underlying
network  reduces routing delays and network link traffic.
Every network node $x$ which corresponds to a leaf in the partition
tree is assigned a {\it Plane Equation} or PE to specify its region in
the whole space.
\subsection{Space Partitioning}
The partition tree is the main data structure in RAQ which
partitions the search space into $n$ regions corresponding to $n$
nodes. Assuming that $r$ is the root of partition tree and representing
the whole search space, each internal node   divides its region into
two smaller regions using a hyperplane equation.  Although only
leaves in the partition tree represent actual network nodes, each
node in this tree has a corresponding region in the search space.
Every network node $x$ which corresponds to a leaf in the partition
tree is assigned a {\it Plane Equation} or PE to specify its region in
the whole space. Each PE consists of some paired labels which is
defined as $X_{PE} = ((p_{1}, d_{1}),(p_{2}, d_{2}), \cdots,
(p_{r(x)}, {d}_{r(x)}))$. In each label, $r(x)$  presents the
distance of $x$ from the root of the tree and $p_i$ shows the plane
equation that partitions the $i$th region into two regions and $d_i$
determines one side of the plane $p_i$ (left or right). Every leaf node in the RAQ stores its own PE as well as the PE of its links.  Figure~\ref{fig:fig1} (left) shows partitioning of 2-dimension search space.
We use ``+" and ``-" in the PE of nodes to determine one side of the plane (left or right).
\subsection{Network Links}
Every node has some links to other nodes of RAQNet overlay. Each link
is  the addressing information of the target node which is its
IP address and its PE. Connection rules in RAQ are based on the partition
tree. Consider node $x$ and its PE, $x$ has link to one of nodes
in each of these sets: $[((p_{1}, \bar{d}_{1}))]$,
$ [((p_{1},
d_{1}), (p_{2}, \bar{d}_{2}))]$,$\cdots$, $[((p_{1}, d_{1}), (p_{2},
d_{2}), \cdots,$ $ (p_{r(x)},$ $\bar{d}_{r(x)}))]$, where $\bar{d}_{i}$
is the opposite side of $d_{i}$. It is easy to show
 that each node has links to $O(\log n)$ nodes in
RAQ.
\subsection{Query Routing }
\begin{figure}[ht]
\includegraphics[scale=0.4]{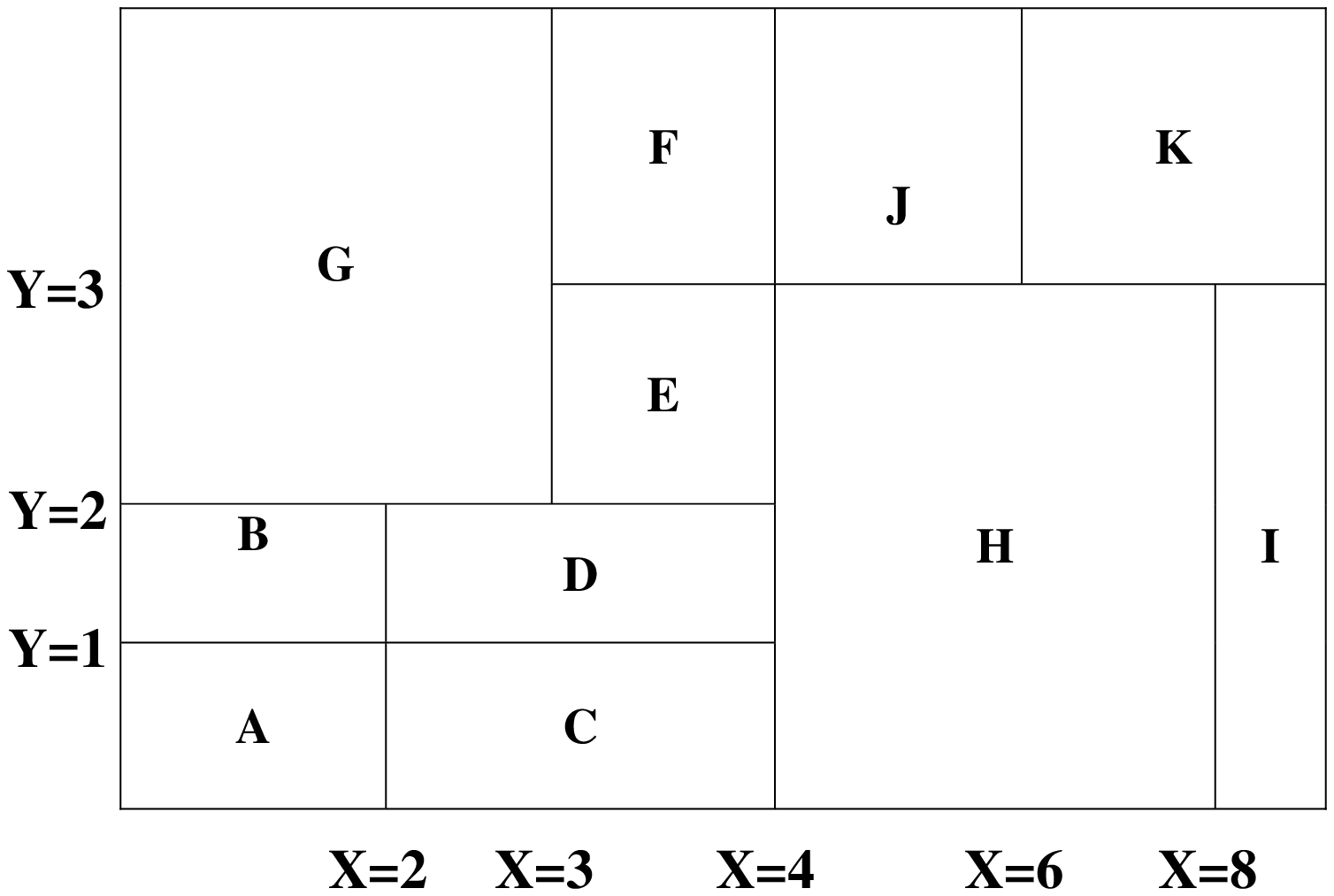}
\hspace{.01cm}
\includegraphics[scale=0.4]{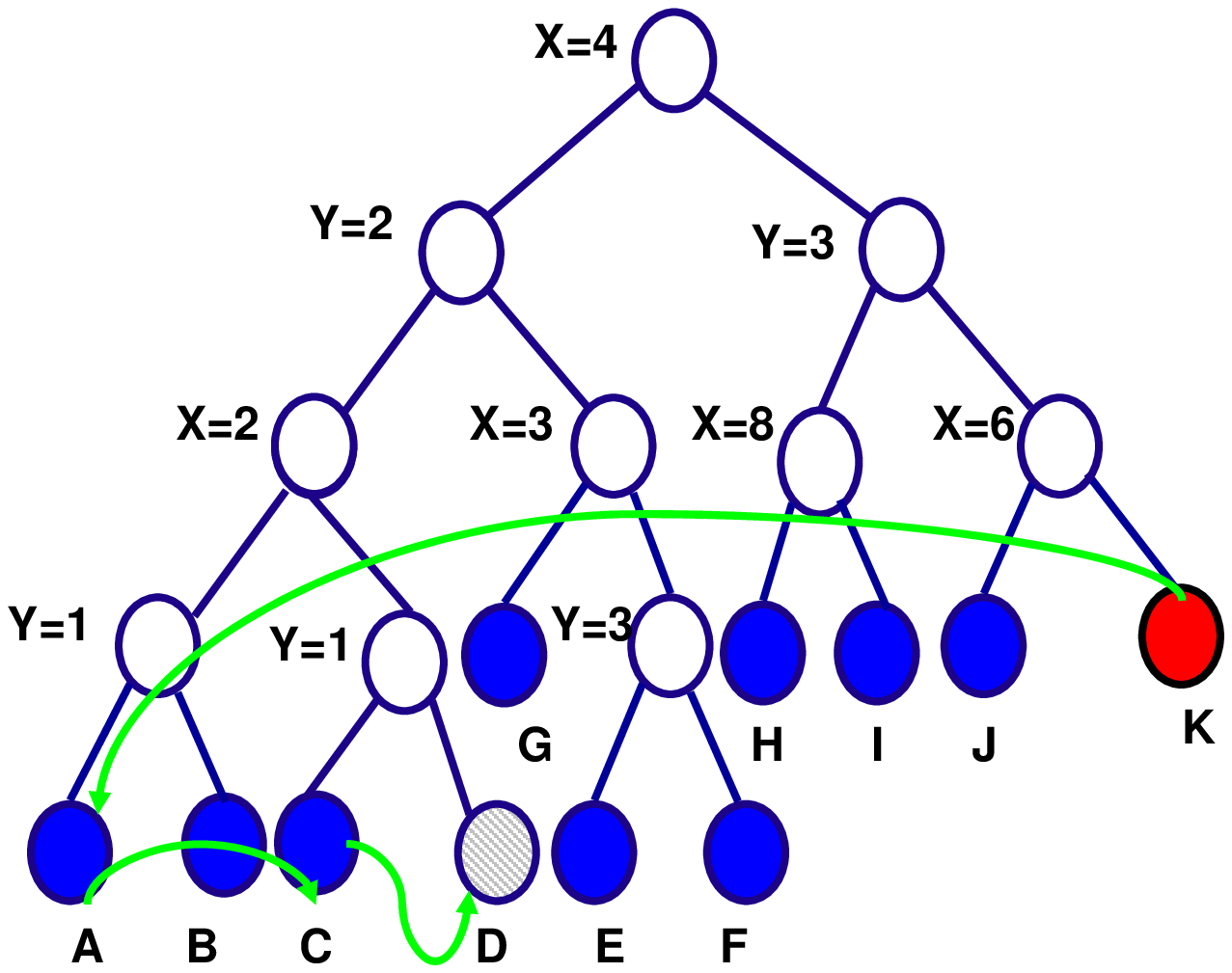}
\caption{Right: 2-dimension search space, Left: Routing a query from node $k$ with PE $[(x=4,+), (y=3,+),(x=6,+)]$ to destination point $(2.5,1.5)$.}
\label{fig:fig1}
\end{figure}

At each routing step, a node usually sends a query to another
node that its PE shares at least one label longer with the destination
point  than the prefix with the present node's PE. If no such node is
known, the query is sent to a node with PE closer to the
destination point and shares a prefix with the destination point
having  the same length. If there is no such node, the present node will be considered as the closest node to the destination point. In figure~\ref{fig:fig1} (right), it shows routing a query from node $k$ to destination point $(2.5,1.5)$.
\subsection{Neighbor Selection Based on Topology Awareness}\label{neighbor selection}
RAQNet seeks to
exploit topology awareness from its underlying network in order to
fill its routing table rows effectively.
 Hence, any node with the
required prefix in PE is used to fill an entry.
Topology aware
neighbor selection selects the collection of \emph{close} nodes among nodes with PE having the required prefix. Topology
awareness relies on a  proximity metric that indicates the
``distance" between any given pair of nodes. The choice of a
proximity metric depends on the desired quality of overlays (e.g., low delay, high bandwidth). Our proximity metric in
RAQNet overlay is round trip delay, estimated by the \emph{PIC} mechanism~\cite{PIC}.
\subsection {In-degree of  Nodes}\label{nodelinks}
Maintaining topology-aware property  effects on the in-degree of  nodes. In the RAQ data structure  or in overlay networks like CAN~\cite{CAN}, and Chord~\cite{chord} that do not consider topology awareness, the in-degree of  node $x$, the number of routing table entries of overlay nodes refers to node $x$, should be balanced across all nodes because
the network links between nodes are selected uniformly according to their connection rules. This feature is appealing because it balances the query routing.
When routing table entries are filled with the \emph{close} nodes, it  causes that the distribution of  in-degree of nodes is affected by the topology of underlying network. Therefore, there is a tradeoff between providing topology awareness and balancing the in-degree  of nodes in the routing algorithm. In the section \ref{Loadbalancelinks}, we quantify the degree of imbalance caused by topology-aware property and we propose a reactive approach to handle it.
\section{Topology Aware Load Balancing}\label{topologyloadsec}
In this section, we present the concept of virtual servers  and then introduce  our topology-aware load balancing algorithm.
\subsection{Virtual Servers in RAQNet}
The concept of virtual servers was first introduced in Chord~\cite{chord} to improve load balancing of overlay nodes. A virtual server looks similar to a single node which is accountable for a region of the search space. Several virtual servers can be hosted by a physical node.
Therefore, any physical node possesses  noncontiguous regions of the search space.
Each virtual server owns its routing tables and stores data items with IDs falling into
its accountable region.

Any virtual server causes definite amount of load. For instance, serving queries which fall into  accountable region of a virtual server
 generates  load. Whenever a node becomes overloaded, it transfers portion of its load to some lightly loaded nodes to become light in which the basic unit of load transfer is virtual servers~\cite{loadbalancesp,loadbalancedsp}. Therefore,  transferring virtual servers from heavy nodes to light nodes causes load balance. The transfer of a virtual server is implemented as a departure operation comes before a join operation, all overlays provide these operations.

When a node $x$ leaves the overlay, its  regions  are taken by other nodes which have contiguous regions with virtual servers of node $x$~\cite{RAQNet}. If a virtual server $v$ leaves the overlay, its responsible region is taken by another virtual server  which has contiguous region with $v$. If there is no such virtual server, region of $v$ is taken by a virtual server with  PE closer to $v$.
 In the same way, When a new node $x$  joins the overlay, it chooses $numVS$ (number of virtual servers) random point $X$ in the search space and sends
its join request.

One disadvantage of using virtual servers is that
any overlay node maintains $numVS$ routing states for its virtual servers.
Our experimental results show that our approach reaches good load  balance when each node has $numVS=\log N$ virtual servers.
In our belief, this overhead can be reasonable. One of the main benefits of using virtual servers is that no overlay modification is needed to perform load balancing algorithms.
\subsection { Topology Aware Load Balancing Algorithm}
The virtual server reassignments are done with regards to topology information of underlying network.
As we described in RAQNet~\cite{RAQNet},
routing tables of  node are filled by close nodes according to proximity metric of overlay network during node join.
 RAQNet overlay nodes inherently maintains information of \emph{close} nodes. Our load balancing algorithm use this information and the {\emph PIC}  mechanism~\cite{PIC} to predicate network distance(i.e., round-trip delay or network hops) between  light  and heavy nodes during virtual server reassignments, described in section~\ref{Reassigning}. The {\emph PIC} mechanism  predicts the distance between two overlay nodes only by having their practical internet coordinates.

We have two assumptions in our load balancing approach:
we attempt to optimize  only one bottleneck resource and we suppose that the load on a virtual server is stable  while carrying out our load balancing algorithm.

These are some definitions we use to explain our approach:
\begin{description}\label{description}
    \item []  {\bf Utilization}:   $ u_i$ is the ratio of node $i$ load to its capacity; $u_i = \frac{l_i}{c_i}$. The $l_i$ shows load of node $i$ at a definite time and each node $i$ has a  capacity $c_i$ which may represent available storage, processor speed, or bandwidth.
	\item [] {\bf Neighborhood Utilization}:  neighborhood utilization of node $i$ is defined as $Neighutil_i = \frac{\sum_{i=1} ^ l Load_i}{\sum_{i=1} ^ l Capacity_i}$, where $s$ is a set of \emph{close} nodes
which announce their load information to node $i$ and  $l$ is the number of nodes in set S.
	\item[] {\bf Load transfer cost}: Load transfer cost  is defined as $LTC=\sum_{i=1}^ n Load_i*Dist_i$,
where
$Dist_i$ indicates the network distance to transfer load of node $i$. The amount of transferred load for node $i$ is shown by $Load_i$.
  \end{description}
\subsection {System Overview}
Our load balancing approach  is not limited to a special kind of resource (e.g., storage, bandwidth, or CPU) and  includes three steps:
\begin{enumerate}
    \item Neighborhood Load Information Collection. All nodes collect load and capacity information of
nodes that exist in their routing table entries with network distance  less than the one desired by
the node for transferring load, we called $DesiredVal$.
	\item Node Categorization. According to  the load and capacity of neighborhood nodes,
all nodes categorize themselves into heavy or light nodes.
	\item  Virtual Server Reassignments. After collecting  neighborhood  load information, computing the neighborhood utilization of overlay nodes, categorizing nodes, and finding proper  light nodes which
are \emph{close} according to $QLB$, a compromised parameter between the
quality of load balancing and load transfer cost, heavy nodes transfer their  virtual servers to light nodes.
\end{enumerate}
\subsection{Neighborhood Load Information Collection} \label{Neighborhood}
We use the procedure in figure~\ref{ProbingLoad} to choose nodes and collect their neighborhood load information.
Before calling this procedure, each node which wants to decrease its load determines the maximum desired network hops  or round trip time($DesiredVal$) to transfer its extra load.
In fugure~\ref{ProbingLoad}, The $Dist_{ node-node_{r}}$ calculates the distance between $node$ and the node in rth row of its routing table
 by the \emph{PIC} mechanism.

We use a  restricted flooding schema to collect neighborhood load information. The flooding schema with a few Time-to-Live (TTL) hops have been presented by S. Jiang et al.~\cite{light-flood}. They have shown that is extremely effective and generates  few excess messages.
Regularly, each node $i$ sends a probing message including the origin address information(IP), its practical internet coordinate(for estimating network distance), the $DesiredVal$ value and a  TTL value to some nodes that exist in its routing table entries, with network distance to node $i$ less than the $DesiredVal$.
\begin{figure}
\scriptsize
{
(Int $DesiredVal$, Node $node$)
{ \bf Procedure Probe-load}
\begin{enumerate}
\itemsep=0mm
\item {   rowNum $=$ getNumberofRoutingTableRows($node$) }
\item {$counter=0$}
\item {\bf if} {(TTL $ \neq 0$)}
\item \ \ { \bf for} (r=0; r$ < $rowNum; counter++)
\item \ \ \ \  { \bf if }($Dist_{ node-node_{r}}$ $ < Desired Val$)
\item \ \ \ \ \ \ SendProbingLoadMsg to $node_{r}$ (TTL, $DesiredVal$,

\ \ \ \ \ \  Coordinate of $node$)
\item \ \ \ \  { \bf end if }
\item \ \  { \bf end for}
\item  { \bf end if }
\end{enumerate}
}
\normalsize
\caption{Probing neighborhood load information. \label{ProbingLoad}}
\end{figure}
\begin{figure}
\scriptsize
{
{ \bf Procedure Find-LightNode($Virtual Server$ Candidate-VS)}
\begin{enumerate}
\itemsep=0mm
\item {   $CanNode = nil$; }

 // $CanNode$ is a candidate  node to recieve virtaul servers.
\item  {\bf for} (each Node $j$ in  $NLIS_i$)

//$NLIS_i$ is  neighborhood load information set of node $i$.
\item \ \ \ \ \  $T_j=Neighutil_i * C_j$

//  $Neighutil_i$ is neighborhood utilization of node $i$.
\item \ \ \ \ \   { \bf if }($Load_j$ + $ Load_{Candidate-VS} < T_j$ )
\item \ \ \ \ \   \ \ \ \ \ \ { \bf if} ($CanNode== nil$)
\item \ \ \ \ \   \ \ \ \  \ \ \ \ \ \ $CanNode= Node_j$
\item \ \ \ \ \   \ \ \ \  \ \ \ \ \ \ $Dist_{ReNode}=Dist_{i-heavyNode}$
\item \ \ \ \ \   \ \ \ \  \ \ \ \ \ \ $U_{CanNode}= U_j$
\item \ \ \ \ \   \ \ \ \  \ \ {\bf else}
\item \ \ \ \ \   \ \ \ \  \ \ \ \ \ \ $Dist_i$=$Dist_{i-heavyNode}$

 \ \ \ \ \   \ \ \ \  \ \ \ \ \ \ //$QLB$ is a tradeoff parameter between the quality of

 \ \ \ \ \   \ \ \ \  \ \ \ \ \ \ //load balancing and load transfer cost
\item \ \ \ \ \   \ \ \ \  \ \ \ \ \ \ { \bf if} ($Absolute  (Dist_{j-heavyNode} -$

 \ \ \ \ \   \ \ \ \  \ \ \ \ \ \ $Dist_{CanNode-heavyNode})\leq  QLB$)
\item \ \ \ \ \   \ \ \ \  \ \ \ \ \ \ \ \ \ \  { \bf if} $(U_j<U_{CanNode})$

\ \ \ \ \   \ \ \ \  \ \ \ \ \ \ \ \ \ \  // prioritize to quality of load balancing.
\item \ \ \ \ \   \ \ \ \  \ \ \ \ \ \ \ \ \ \  \ \ \ \ $CanNode= Node_j$
\item \ \ \ \ \ \  \ \ \ \ \ \ \ \ \ \ \ \ \ \ \ \ \ \ \ \ \ \ $LTC_{ReNode}=LTC_i$
\item \ \ \ \ \   \ \ \ \  \ \ \ \ \ \ \ \ \ \ { \bf end if}
\item \ \ \ \ \   \ \ \ \  \ \ \ \ \ \   { \bf else}
\item \ \ \ \ \   \ \ \ \  \ \ \ \ \ \ \ \ \ \ { \bf if} $(Dist_{j-heavyNode}< Dist_{CanNode-heavyNode})$
\item \ \ \ \ \ \  \ \ \ \ \ \ \ \ \ \ \ \ \ \   \ \ \ \ \ \ \ \ LTC-ReNode$=LTC_i$

\ \ \ \ \   \ \ \ \  \ \ \ \ \ \ \ \ \ \ // prioritize to network distance.
\item \ \ \ \ \   \ \ \ \  \ \ \ \ \ \ \ \ \ \ \ \ \ \ $CanNode= Node_j$
\item \ \ \ \ \   \ \ \ \  \ \ \ \ \ \ \ \ \ \ { \bf end if}
\item \ \ \ \ \   \ \ \ \  \ \ \ \ \ \   { \bf end if}
\item \ \ \ \ \   \ \ \ \ \ \ { \bf end if}
\item \ \ \ \  \  { \bf end if  }
\item  {\bf end for}
\item { \bf return} CanNode
\end{enumerate}
}
\normalsize
\caption{Finding light nodes to receive a virtual server. \label{Finding}}
\end{figure}
 \begin{figure}[ht]
\scriptsize
{
{ \bf Procedure Reassign-VirtualServer}
\begin{enumerate}
\itemsep=0mm
\item {   Node $i$ calculates it's neighborhood utilization }
\item $T_i=(Neighutil_i+\varepsilon) * C_i$

 // $T_i$ is target load of node $i$.
\item    { \bf if }($L_i \leq T_i$) \ \ \ \ \ \
\item  \ \ \ \ { \bf return; } // Node $i$ is a light node.
\item   { \bf end if }
\item   Candidate-VS = Node $i$ chooses one of its VS to leave.

 /* VS is abbreviation  of Virtual server*/
\item  Receiving-Node =  Find-LightNode($Candidate-VS$)
\item  {\bf if} (Receiving-Node!= null)
\item  \ \ \ \  Transfer Candidate-VS to Receiving-Node
\item  {\bf end if}
\end{enumerate}
}
\normalsize
\caption{ Reassigning virtual servers. \label{Reassignment}}
\end{figure}
Node $j$ which receives a probing message replies to the origin node $i$ with its address information, current load, capacity and its practical internet coordinate. Then,
The TTL value is decreased by 1 and if the updated TTL value does not reach to 0,
it resends the received probing message  to its routing table entries with network distance to origin
node $i$ less than the $DesiredVal$.
When origin node $i$ receives the replied probing messages, it
 computes the round trip time to the responding nodes by using the practical internet coordinate of responding nodes and the\emph{ PIC} mechanism. After that, node $i$ stores this information to its neighborhood load information set($NLIS_i$). The member count of this set can be represented as following:
MemberCount($NLIS_i$) =
$\sum_{j=1}^{TTL} numVS^j = \frac {numVS*(numVS^{TTL}-1)} {numVS-1}=$

$O(numVS^{TTL})$.
In this formula, $numVS$ represents the number of virtual servers per overlay node.
Based on our experimental evaluations in section \ref{Exresult}, our approach reaches a fine load balance if $numVS$ is $O(\log n)$ and TTL is 2.
So, The member count of $NLIS_i$ is $O(\log^2 n)$ in the worst-case.
We consider the worst-case to be sending the probing messages to all
 nodes in the routing table entries and it happens only if we assign the biggest possible value to the
 $DesiredVal$.
\subsection{Node Categorization and Virtual Server Reassignments}\label{Reassigning}
Whenever  a neighborhood load information set is become ready, each node $i$ knows the load and capacity of neighborhood nodes and then calculates its neighborhood utilization, $Neighutil_i$, and its target load, $T_i$.
After that, if its current load, $L_i$, is bigger than its target load, $T_i$,  it marks itself as a heavy node , then it chooses one of its virtual servers to leave node $i$ and makes it light. Finding a proper  virtual server takes $O( numVS)$ time.

We use procedure  $Reassign-VirtualServer$ to reassign virtual servers. This process  is described in
 figure~\ref{Reassignment}. $\varepsilon$ is a parameter for a tradeoff between the amount of load transferred and the quality of load balancing. $\varepsilon$ ideally is 0.
Calculating the best reassignments is equivalent to minimize maximum node utilization  problem and
is NP-complete~\cite{NP-Complete}. So, it is impossible to reassign virtual servers across nodes perfectly but it can be solved by an approximate algorithm.

We use the procedure in  figure~\ref{Finding} to find a candidate node, $CanNode$, to recieve virtual servers. Each heavy node $i$ finds a proper light node from its neighborhood load information set, $NLIS_i$.
In figure~\ref{Finding},
utilization of node $j$ is presented with $U_j$, and $T_j$ indicates target
load of node $j$.
$ Load_{Candidate-VS}$ shows the load of candidate virtual server of node $i$, supposed to leave node $i$.
 where $CanNode$ is a candidate node to receive .
 The network distance between node $i$  and $ReNode$ is shown by $Dist_{i-ReNode}$.
In this procedure, it is certain that load of new virtual servers do not cause $CanNode$ to become heavy, shown in line $4$. Additionally, The $Dist_{j-CanNode}$ calculates the distance between node $j$ and node $CanNode$  according to their practical internet coordinates by the \emph{PIC} mechanism.
Our approach provides a different kind of tradeoff between the quality of load balancing and load
transfer cost, based on parameter QLB. A heavy node chooses a light node with the smallest utilization and network distance less than the $QLB$. Otherwise it tries to find the closest light node with network distance more than the $QLB$, shown from line $9$ to $16$.

The worst-case running time (to be sending the probing message to all nodes in the routing table entries) of procedure \emph{Find-LightNode} is $O(numVS^{TTL})$, where the average number of virtual servers is shown by $numVS$. Based on our experimental results, whenever $numVS$, $TTL$, $QLB$ and $DesiredVal$  are equal to
$O(\log N)$, 2, $130$ and $400$(based on GT-ITM\cite{GT-ITM}) , our algorithm achieves good load balance. Additionally, the  worst-case running
 time of our proposed approach is $O(\log N+\log^2 N)$. So that it  does well in term of scalability.
\subsection {Synchronization between Light and Heavy Nodes}
All overlay nodes  collect load information and reassign virtual servers concurrently. Therefore, some virtual servers may be sent to a light node from different heavy nodes, which causes a light node to become overloaded. Hence, before sending virtual servers, a heavy node send a $synch$  message to a light node. If its load is not changed, it acknowledges the heavy node and does not acknowledge to others. After a distinct interval,
 if a heavy node does not receive any acknowledge message from a light node,  it chooses
 another node to transfer its extra load.

\section{Experimental Results}\label{Exresult}
We present experimental results which evaluates our load balance approach in RAQNet overlay network.  The results were achieved using a RAQNet overlay with 4096 nodes  running on an Internet topology model.
We assume that $f$ is a fraction of the search space which belongs to a virtual server that is exponentially  distributed. Also, $\mu$ and $\sigma$ show the mean and the standard deviation of total load on RAQNet overlay. We use Gaussian distribution with mean $\mu f$ and the standard deviation $\sigma \sqrt{f}$\cite{loadbalancesp} for the load on virtual servers. We also use $Gnutella-like$
capacity for capacity of nodes. Consequently, 20 percent,
45 percent, 30 percent, 4.9 percent, and 0.1 percent of node capacity  is  1, 10, 100, 1000, 10000.

Our experiments run on a simulated network topology which was generated by the Georgia Tech transit-stub network topology model~\cite{GT-ITM}. Ts4k-small includes 4 transit domains each with 4 transit
nodes, 5 stub domains connected to each transit node, and 55 nodes in each stub domain on average.
\subsection{The Effect of  Load balancing Parameters }
\begin{figure}[h]
\includegraphics[scale=0.4]{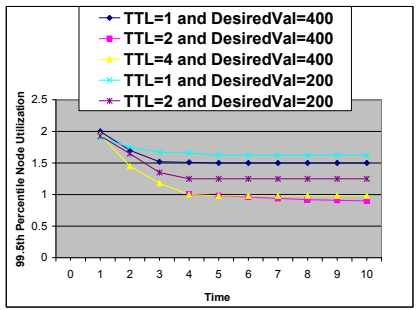}
\caption{The effect of various load balancing parameters   on node utilization.}
\label{fig:TTL}
\end{figure}

We assign $DesiredVal=400$ and $QLB=130$(based on GT-ITM\cite{GT-ITM}) while  performing our topology-aware load balancing algorithm.
Figure~\ref {fig:TTL} shows that the TTL value  affects on  node utilization. It  improves the quality of load balancing while TTL value  changes from 1 to 2 because there exist more alternative light  nodes  in  neighborhood load information set, as we said in section \ref{Neighborhood}.

Increasing $QLB$ , the algorithm gives priority to the quality of load balancing and it
ignores the load transfer cost. When  we assign the biggest possible amount to $DesiredVal$ and $QLB$,
our approach performs topology unaware load balancing.
By decreasing  the value of $DesiredVal$, fewer nodes will report their load information to requesting nodes and the quality of load balancing will be decreased in overlay network.
When TTL value is increased up to 4 or even more, the quality of load balancing will be decreased
surprisingly.
This is because
it may ignore the nodes  with small utilization and far network distance. However,   the values of $QLB$ and $desiredVal$  will affect the quality of load balancing. Thus, each overlay node can compromise between load transfer cost and its utilization.
We  separately compute the LTC (load transfer cost), defined in section\ref{description}, with and without considering topology awareness. Then, we calculate
$Benefit$ $= \frac{LTC_{Without topology}-LTC_{topology}}{LTC_{Without topology}}$. $Benefit$ is $43\%$ in GT-ITM topology model. Therefore, the network bandwidth is saved greatly in our approach.
\subsection{Topology Aware Load Balancing}
\begin{figure}[h]
\includegraphics[scale=0.4]{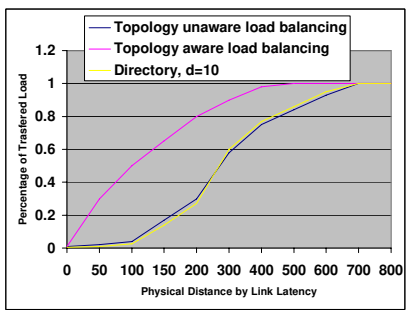}
\caption{Cumulative distribution of transferred load  in GT-ITM topology.}
\label{LoadBalancing}
\end{figure}

In this section, we show the  effect of topology awareness on load balancing. In figure~\ref{LoadBalancing}, cumulative distribution of transferred load is illustrated. It shows that 50\%  of load  is transferred
via the network distance with average link  latency of 100 in GT-ITM topology. Also, more than
80\% of load traverses the links with total average latency of 200. In contrast,
regardless of topology awareness, the 50\% of load is transferred having average of about 280 of link latency. Consequently, extra load of heavily loaded nodes is transferred among \emph{close} nodes and it imposes less
traffic to underlying network. Thus, Topology aware load balancing saves bandwidth considerably.
Moreover, this algorithm converges quickly because it chooses nodes from  \emph{close} groups which are physically  \emph{close} together and therefore reduces the cost of transferring load.
In figure~\ref{LoadBalancing} the load balancing algorithm, suggested by Godfrey et al~\cite{loadbalancedsp}, is indicated
by "directory" line. It is obvious that their method is similar to topology unaware load balancing.

\begin{figure}[h]
\includegraphics[scale=0.3]{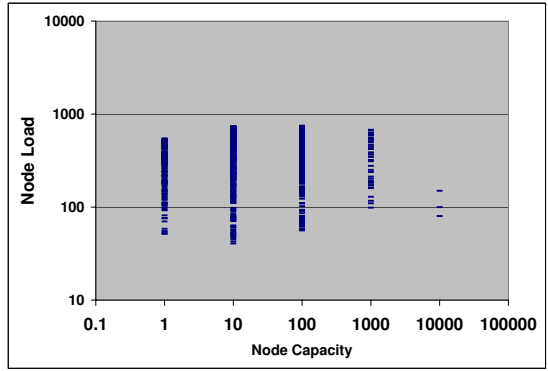}
\caption{The scatter plots of load and capacities before load balancing. }
\label{fig:capacity1}
\end{figure}

\begin{figure}[ht]
\includegraphics[scale=0.3]{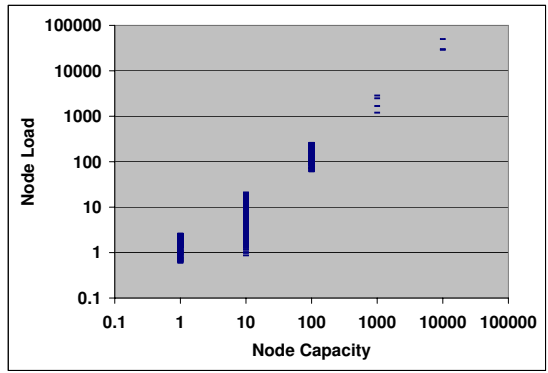}
\caption{The result of load balancing algorithm. }
\label{fig:capacity2}
\end{figure}

The scatter plots of load for the Gaussian
distribution are shown  in figure~\ref{fig:capacity1} and figure~\ref{fig:capacity2}
 and we use a $Gnutella-like$ capacity in our  node capacity model. Our load balancing approach helps to  rearrange a bad load distribution into an acceptable arrangement and eventually each overlay node will have the load proportional to its capacity.

\subsection{Balancing  the In-degree of  Nodes}\label{Loadbalancelinks}
\begin{figure}[t]
\hspace{0.5cm}
\includegraphics[scale=0.3]{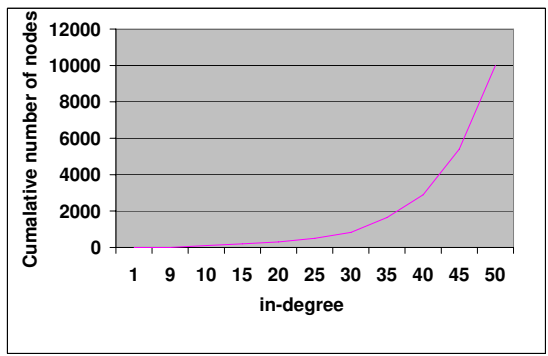}
\caption{The in-degree distribution of 10000 nodes. }
\label{fig:load1}
\end{figure}
Figure~\ref{fig:load1} illustrates the cumulative in-degree distribution  of 10000 RAQNet nodes,
based on GT-ITM topology. The result shows that this distribution is not completely balanced.
We also notice that this imbalance is quite remarkable in the first rows of routing tables because these
rows are filled with  \emph{close} nodes that their PE matches with the first labels of PE in local node. Obviously, choosing  \emph{close} nodes will cause  in-degree imbalance.

We deal with this problem reactively.
If one of the nodes with a high in-degree has the heavy  workload, it sends a $heavyLoad$ message to
its neighbors by the departure links, as we explained in RAQNet. The nodes which receive a $heavyLoad$
message mark the corresponding entry in their routing tables and try to find an alternative node by triggering the routing table maintenance procedure which is described in RAQNet. Because the most remarkable entries happen to be at the top rows of routing tables, replacing nodes will not increase the  distance traversed by messages.
We believe that the in-degree imbalance as a consequence of topology awareness does not lead in a remarkable problem.
\section{Conclusion}\label{Concsec}
This paper presents a simple distributed  load balancing algorithm with topology-aware property for structured P2P overlay networks.
In our approach, each node collects neighborhood load information from  \emph{close}  nodes and then it reassigns  its own extra load according to topology of underlying network. Consequently, it provides rapid convergence on load balance and reduces the load transfer cost.
Our parametric algorithm  increases the quality of load balancing among \emph{close} nodes and also  provides
a different kind of tradeoff between the quality of load balancing and load transfer cost. The experimental results show that this approach is effective and  considerably saves network bandwidth.

Additionally, We conclude that the in-degree imbalance of nodes does not cause a significant problem
in topology aware overlays and also we propose a reactive solution to deal with it.

We plan to enhance our load balancing approach to adapt in a dynamic system. Moreover, as a future improvement to our approach, imposing other constraints (e.g. utilization of nodes) during  collecting the load information of nodes, may be considerably helpful.


\end{document}